\documentclass[onefignum,onetabnum]{siamart171218}
\usepackage[utf8]{inputenc}
\usepackage{ulem,bm}
\usepackage{caption}
\usepackage{amsmath}
\usepackage{changepage}
\title{Tipping Cycles}
\author{Michael A.S. Thorne  \\ British Antarctic Survey} 
\date{}

\begin{document}

\maketitle

\textbf{Ecological systems are studied using many different approaches and mathematical tools. One approach, based on the Jacobian of Lotka-\\Volterra type models, has been a staple of mathematical ecology for years, leading to many ideas such as on questions of system stability. Instability in such methods is determined by the presence of an eigenvalue of the community matrix lying in the right half plane. The coefficients of the characteristic polynomial derived from community matrices contain information related to the specific matrix elements that play a greater destabilising role. Yet the destabilising circuits, or cycles, constructed by multiplying these elements together, form only a subset of all the feedback loops comprising a given system. This paper looks at the destabilising feedback loops in predator-prey, mutualistic and competitive systems in terms of sets of the matrix elements to explore how sign structure affects how the elements contribute to instability. This leads to quite rich combinatorial structure among the destabilising cycle sets as set size grows within the coefficients of the characteristic polynomial.}\\

Within mathematical ecology, one approach [2] to represent predator-prey, mutualistic and competitive systems is through the community matrix, a real-valued square matrix representing the linearisation of Lotka-Volterra type dynamic equations ([4],[8]). Each of the forms is represented through differing sign conventions as can be seen in the following $3 \times 3$ examples: 

\vspace{.2in}
\begin{center}
\begin{equation*}
\begin{tabular}{l c c c}
    & Predator-Prey & Mutualistic & Competitive \\  
    & 
    $\begin{bmatrix}  
   -a & b & c \\ -d & -e & f \\ -g  & -h & -k 
   \end{bmatrix} $ &
   $\begin{bmatrix}  
   -a & b & c \\ d & -e & f \\ g & h & -k 
   \end{bmatrix}$  & 
   $\begin{bmatrix}  
   -a & -b & -c \\ -d & -e & -f \\ -g & -h & -k 
    \end{bmatrix}$. 
     \medskip \\
 \end{tabular} 
\end{equation*}
\end{center} 
\vspace{.2in}

Stability of a community matrix is determined by whether all of the eigenvalues lie in the left half of the complex plane. The characteristic polynomial of an $n \times n$ community matrix is an $n^{th}$-order monic polynomial (or made so by change of sign), whose roots are the eigenvalues of the system. A necessary condition for the roots of a polynomial to all lie in the left half plane, and therefore for the system to be stable, is for all of its coefficients to be positive. However, this condition is not sufficient. The Routh-Hurwitz stability criteria ([3],[6]) provides a further, and sufficient, condition. The following equation is an example of a polynomial satisfying the necessary condition of positive coefficients,

\begin{equation}
     x^4+2x^3+3x^2+4x+5 = 0, 
\label{eq:rh}
\end{equation}   
yet whose Routh table (Table 1) reflects the existence of two roots in the right half plane and is therefore not Routh-Hurwitz stable.

\begin{table}[!h]
\centering
\begin{tabular}{c| c c c c}
	$x^4$  & 1 & 3 & 5 \\
	$x^3$   & 2 & 4 & \\
	$x^2$  & 1 & 5  &  \\
	$x^1$ &  -6  &  & \\
	$x^0$  & 5 &  & \\
\end{tabular}
\label{tab:rh}
\caption{\textbf{Table 1} \textit{The Routh table for Eq.~\ref{eq:rh}. The number of sign changes as one moves down the first column indicate the number of roots in the right half plane. In this case, two.}}
\end{table}
 
Polynomials of the example just described can only have roots in the right half plane that are complex conjugate pairs. This was proven by Obrechkoff [5], who showed that there are no solutions to polynomials with positive coefficients that lie on the positive real axis (the right half plane roots of Eq.~\ref{eq:rh} are $0.29 \pm 1.42i$). 

Therefore, any systems that have a real maximal (largest real part) eigenvalue (with imaginary part zero) have their tipping point between stability and instability just where the last coefficient of their characteristic polynomial becomes positive. This point is helpful in considering what contributes to the destabilisation of a system through untangling the elements that make up the feedbacks.

The idea of a tipping point is not very informative without its context, without knowledge of the feedbacks that produce the instability. Ecological systems that are modelled by community matrices have the benefit of the direct relation through the coefficients of the characteristic polynomial which are derived from the minors, or respectively the summations of the n-tuple sets of the eigenvalues [1], of the matrix. These minors may be exhaustively described by the terms consisting of sets of multiplied elements of the community matrix. These terms are the feedback loops of the system, otherwise called circuits, or cycles. Within each coefficient of a characteristic polynomial, one can distinguish the cycles that contribute to the coefficient becoming more positive, and therefore stabilising, or more negative and destabilising. For example, consider the general characteristic polynomial of the $3 \times 3$ predator-prey community matrix described above,

\begin{equation*}
             a_3x^3+a_2x^2+a_1x+a_0,
\end{equation*}
which expands, in terms of its matrix elements as, 

\begin{equation*}
 x^3+(a+e+k)x^2+(bd+ae+cg+fh+ak+ek)x+(ceg+bfg+afh+bdk+aek\textcolor{red}{-cdh}).
\end{equation*}
 
Given that characteristic polynomials are monic, we know that the highest order coefficient, $a_3$, is $1$ and therefore positive, $a_2$ is the negative of the trace, and therefore positive, and as can be seen, $a_1$, the summed combination of 2-tuples, is also positive. This leaves only $a_0$ able to switch between being positive or negative, and therefore able to destabilise the system. For clarity of explanation, we have used the change of sign of the coefficient for consideration of stability, even though this is only a clear indication in systems with a maximal real-valued eigenvalue. But the effect of increasing the positive or negative value of a coefficient on its stability applies to all systems. Within $a_0$, highlighted above and below in red, the only destabilising circuit, or tipping cycle, is $cdh$,

\vspace{.1in}
\begin{equation*}  \left[ \begin{array}{rrr} -a & b & {\color{red}c} \\ {\color{red}-d} & -e & f \\ -g & {\color{red}-h} & -k \end{array} \right].   \end{equation*}
\vspace{.1in}

This means that however large the other elements are, they cannot force the system to become unstable (if it is not already so), only the circuit of c, d and h can do that. Given that there are five other cycles of size three that make up the $a_0$ coefficient, it is helpful to consider a ratio, the coefficient feedback sensitivity, $\tilde{a_i}$, which in this case is 

\begin{equation*}  
\tilde{a_0}=1/6.
\end{equation*}

Table 2 shows the values of $\tilde{a_i}$ for predator-prey, mutualistic and competitive systems up to $8 \times 8$ sized community matrices. \\

\begin{table}
\begin{adjustwidth}{-.8cm}{}
\tiny
\begin{tabular}{c c c c c c c c c c c}
 & $n$ & $a_8$$x^8$ & $a_7$$x^7$ & $a_6$$x^6$ & $a_5$$x^5$ & $a_4$$x^4$ & $a_3$$x^3$ & $a_2$$x^2$ & $a_1$$x$ & $a_0$ \\
   P-P& $2$ &&&& & &  & + & + & + \\
      & $3$ & &&&& & + & + & + & 1/6 \\
      &$4$ & &&&& + & + & + & 4/24 & 8/24 \\
      &$5$ &&&& + & + & + & 10/60 & 40/120 & 52/120 \\
      &$6$ & &&+&+&+ & 20/120 & 120/360 & 312/720 & 344/720 \\
      &$7$ & &+&+&+& 35/210 & 280/840 & 1092/2520 & 2408/5040 &  2488/5040 \\
      &$8$ &+&+&+& 56/336 & 560/1680 & 2912/6720 & 9632/20160 & 19904/40320 & 20096/40320 \\
   C &$2$  &&&&&  && + & + & 1/2 \\
     & $3$  &&&&&& + & + & 3/6 & 3/6 \\
     & $4$  & &&&& + & + & 6/12 & 12/24 & 12/24 \\
     &$5$  &&&& + & + & 10/20 & 30/60  & 60/120  & 60/120 \\
     &$6$  &&&+&+& 15/30 & 60/120 & 180/360 & 360/720 & 360/720 \\
     &$7$ & &+&+& 21/42 & 105/210 & 420/840 & 1260/2520 &  2520/5040 & 2520/5040 \\
    &$8$  &+& + & 28/56 & 168/336 & 840/1680 & 3360/6720 & 10080/20160 & 20160/40320 &  20160/40320 \\
  M &$2$  &&&&&&  & + & + & 1/2 \\
     & $3$   &&&& && + &  + & 3/6  & 5/6 \\
      &$4$   &&&&& + & + & 6/12  & 20/24  & 20/24 \\
      &$5$  &&&& + & + & 10/20 & 50/60  & 100/120  & 84/120 \\
      &$6$   &&&+&+ & 15/30 & 100/120 & 300/360 & 504/720 & 424/720 \\
      &$7$   &&+&+& 21/42 & 175/210 & 700/840 & 1764/2520 & 2968/5040 & 2680/5040 \\
      &$8$  &+&+& 28/56 & 280/336 & 1400/1680 & 4704/6720 & 11872/20160 & 21440/40320 & 20544/40320 \\

\end{tabular}
\end{adjustwidth}{}
\normalfont
\label{tab:cfs}
\caption{\textbf{Table 2} \textit{For the three forms under discussion $\tilde{a_i}$ is presented for systems up $n=8$. By the inherent sign symmetries and proportions it would be reasonable to suggest that, for any size $n$, $\tilde{a_i} < 1/2$ in predator-prey (P-P) systems (but as $n$ increases, $a_0$ asymptotically tends to $1/2$), $\tilde{a_i} \ge 1/2$ in mutualistic (M) systems (the case when $\tilde{a_i} = 1/2$ only when the diagonal terms do not play a role in the cycles contributing to destabilisation), and $\tilde{a_i} = 1/2$ in competitive (C) systems. Coefficients indicated by a $+$ have no negative destabilising cycles and therefore $\tilde{a_i}=0$. The total number of terms (denominator) for each $a_i$ is $n!/i!$. The values of the numerators of the highest predator-prey $\tilde{a_i}$ follow the tetrahedral numbers, while the mutualistic and competitive highest order numerators of the $a_i$ are the triangular numbers. The sequence of tipping cycles of the competitive systems for a given $n$ as $i$ decreases follows the sequence of path polynomials of the complete graph $K_n$ [7].}}
\end{table}

Consider the $4 \times 4$ predator-prey community matrix, 
\begin{equation*}  \left[ \begin{array}{rrrr} -a & b & c & d \\ -e & -f & g & h  \\ -k & -l & -m & p \\ -q & -r & -s & -t \end{array} \right],   \end{equation*}
then the coefficient $a_1$, consisting of 3-cycles ($n-i$) is, \\

%\begin{multiline}
\hspace{.5in} cfk+bgk+agl+bem+afm+dfq+bhq+dmq+cpq+ahr+hmr+gpr+ \\  \hspace*{1.3in} aps+fps+bet+aft+ckt+glt+amt+fmt\textcolor{red}{-cel-der-dks-hls}.\\
%\end{multiline}\\

We define the tipping cycle set of $a_1$ as $\hat{a_1}=\{cel,der,dks,hls\}$.
These four negative cycles ($\tilde{a_1}=4/24$), that if strong enough, could tip the $a_1$ coefficient from positive to negative, destabilising the system, consist of only eight distinct matrix elements, $c,d,e,h,k,l,r$ and $s$, with each element included in a differing number of cycles. That is, each element has a specific weight. If we construct a weighted matrix of the elements of this tipping cycle set ($[\hat{a_1}]$) one can clearly see the patterning of the key elements that play a role in destabilisation,

\begin{equation*}
[\hat{a_1}] = \left[ \begin{array}{rrrr} 0 & 0 & 1 & 2  \\ 2  & 0 & 0 & 1 \\ 1 & 2 & 0 & 0 \\ 0&1&2&0  \end{array} \right]. 
\end{equation*}

Applied to all the different forms, we can see that the sign structure of the matrix has an effect on the different elements in their role in the destabilising cycles, 

\vspace{.2in}
\begin{center}
$[\hat{a_0}] = $
\begin{tabular}{l c c c}
    & Predator-Prey & Competitive & Mutualistic \\  
    & 
    $\begin{bmatrix}  
   1 & 1 & 3 & 3 \\ 3 & 1 & 1 & 3 \\ 3 & 3 & 1 & 1 \\ 1 & 3 & 3 & 1 
   \end{bmatrix} $ &
   $\begin{bmatrix}  
   3 & 3 & 3 & 3 \\ 3 & 3 & 3 & 3 \\ 3 & 3 & 3 & 3 \\ 3 & 3 & 3 & 3 
    \end{bmatrix}$ &
           $\begin{bmatrix}  
   5 & 5 & 5 & 5 \\ 5 & 5 & 5 & 5 \\ 5 & 5 & 5 & 5 \\ 5 & 5 & 5 & 5 
   \end{bmatrix}$   

     \medskip \\
\end{tabular} \\
$[\hat{a_1}] = $
\begin{tabular}{l c c c}
      & 
    $\begin{bmatrix}  
   0 & 0 & 1 & 2 \\ 2 & 0 & 0 & 1 \\ 1 & 2 & 0 & 0 \\ 0 & 1 & 2 & 0 
   \end{bmatrix} $ &
   $\begin{bmatrix}  
   3 & 2 & 2 & 2 \\ 2 & 3 & 2 & 2 \\ 2 & 2 & 3 & 2 \\ 2 & 2 & 2 & 3 
    \end{bmatrix}$ &
   $\begin{bmatrix}  
   3 & 4 & 4 & 4 \\ 4 & 3 & 4 & 4 \\ 4 & 4 & 3 & 4 \\ 4 & 4 & 4 & 3 
   \end{bmatrix}$  

     \medskip \\
\end{tabular} \\
$[\hat{a_2}] = $
\begin{tabular}{l c c c}
    &
     $\begin{bmatrix}  
   0 & 0 & 0 & 0 \\ 0 & 0 & 0 & 0 \\ 0 & 0 & 0 & 0 \\ 0 & 0 & 0 & 0 
   \end{bmatrix} $ &
   $\begin{bmatrix}  
   0 & 1 & 1 & 1 \\ 1 & 0 & 1 & 1 \\ 1 & 1 & 0 & 1 \\ 1 & 1 & 1 & 0 
    \end{bmatrix}$ &
   $\begin{bmatrix}  
   0 & 1 & 1 & 1 \\ 1 & 0 & 1 & 1 \\ 1 & 1 & 0 & 1 \\ 1 & 1 & 1 & 0 
   \end{bmatrix}$   
     \medskip \\
 \end{tabular}
\end{center} 
\vspace{.2in}

In addition, the total weighting of the elements in their role in destabilisation across all the coefficients are a summation of the individual coefficient tipping cycle set weightings, 

\vspace{.2in}
\begin{center}
\begin{equation*}
\sum_{i=0}^n [\hat{a_i}] =
\begin{tabular}{l c c c}
    & Predator-Prey & Competitive & Mutualistic\\  
    & 
    $\begin{bmatrix}  
   1 & 1 & 4 & 5 \\ 5 & 1 & 1 & 4 \\ 4 & 5 & 1 & 1 \\ 1 & 4 & 5 & 1 
   \end{bmatrix} $ &
   $\begin{bmatrix}  
   6 & 6 & 6 & 6 \\ 6 & 6 & 6 & 6 \\ 6 & 6 & 6 & 6 \\ 6 & 6 & 6 & 6 
    \end{bmatrix}$ &
   $\begin{bmatrix}  
   8 & 10 & 10 & 10 \\ 10 & 8 & 10 & 10 \\ 10 & 10 & 8 & 10 \\ 10 & 10 & 10 & 8 
   \end{bmatrix}.$  

     \medskip \\
 \end{tabular} 
\end{equation*}
\end{center} 
\vspace{.2in}
Extending the summed weightings to the $5 \times 5$ systems, 

\vspace{.2in}
\begin{center}
\begin{equation*}
\begin{tabular}{l c c c}
    & Predator-Prey & Competitive & Mutualistic\\  
    & 
    $\begin{bmatrix}  
   12 & 11 & 20 & 23 & 24 \\ 24 & 12 & 11 & 20 & 23 \\ 23 & 24 & 12 & 11 & 20 \\ 20 & 23 & 24 & 12 & 11 \\ 11 & 20 & 23 & 24 & 12 
   \end{bmatrix} $ &
   $\begin{bmatrix}  
   30 & 25 & 25 & 25 & 25 \\ 25 & 30 & 25 & 25 & 25 \\ 25 & 25 & 30 & 25 & 25 \\ 25 & 25 & 25 & 30 & 25 \\ 25 & 25 & 25 & 25 & 30 
    \end{bmatrix}$ &
   $\begin{bmatrix}  
   46 & 38 & 38 & 38 & 38 \\ 38 & 46 & 38 & 38 & 38 \\ 38 & 38 & 46 & 38 & 38 \\ 38 & 38 & 38 & 46 & 38 \\  38 & 38 & 38 & 38 & 46
   \end{bmatrix}.$  

     \medskip \\
 \end{tabular} 
\end{equation*}
\end{center} 
\vspace{.2in}

The symmetry seen in the competitive and mutualistic $\sum_{i=0}^n [\hat{a_i}]$ are clearly a consequence of the symmetry in the sign structure of these two forms, while for the predator-prey systems it reflects its lack of sign symmetry. Such bias in the weighting of certain individual elements in the predator-prey case raises the question as to how much stability or instability is an artifact of the structure. For example, in larger systems, positioning of one species over another in the governing equations from which the community matrices are constructed may prove crucial for the outcome. That is, it posits the question of whether the medium may well be the message. \\

If the elements comprising tipping cycles are considered in terms of set inclusion, in which for a given $\hat{a_i}$ set comparability is considered as an antichain, and the set inclusion properties of smaller to larger sized cycles is likewise determined by the collection of elements, then we can describe the inclusion properties of $\hat{a_i}$ as $i$ decreases for a given $n$. To see this, we again look at the $4 \times 4$ predator-prey system, where the 4-cycles are $\hat{a_0}=\{chlq,dgkr,demr,cepr,dfks,bhks,ahls,celt\}$. In terms of the constituent elements, the set of 3-cycles of $\hat{a_1}$ are subsets of some of the 4-cycles of $\hat{a_0}$ (i.e. cel $\in$ celt): 
\begin{equation*}
     \hat{a_1} \subset \hat{a_0}.
\end{equation*}
This property, in which the smaller tipping cycle sets are a subset of the larger sized sets holds across all the forms of matrices for a given $n$: 
\begin{equation*}
     \hat{a_i} \subset \hat{a}_{j, \ \forall j < i}.
\end{equation*}
Yet the way the different forms (i.e. predator-prey, mutualistic and competitive) result in different element weightings is also reflected in the different ways their $\hat{a_i}$ flow through their $a_i$ as $i$ decreases. 

For example, with the $4 \times 4$ predator-prey system, there are four tipping cycles in $\hat{a_1}$ and eight tipping cycles in $\hat{a_0}$. While the four smaller sets of $\hat{a_1}$ are subsets of sets in $\hat{a_0}$, the four remaining sets of $\hat{a_0}$ do not have such an embedding from sets of $\hat{a_1}$, but are new configurations of elements from the $4 \times 4$ system (i.e. $\{cel,der,dks,hls\} \subset \{celt,demr,dfks,ahls\}+\{chlq,dgkr,cepr,bhks\}$).

In the $4 \times 4$ competitive case, the six sets in $\hat{a_2}$ are subsets in two sets each in $\hat{a_1}$, but in this case there are no new configurations. Six of the twelve sets of $\hat{a_0}$ consist of two sets each from $\hat{a_1}$ (these are sets constructed through the possible combination of having two of the four diagonal elements, $\binom{4}{2}$), with six new configurations: 

\vspace{.2in}
\begin{center}
\begin{equation*}
\begin{tabular}{l c c c}
    & $\hat{a_2}$ & $\hat{a_1}$  & $\hat{a_0}$ \\  
     \\
    & 
    $\left\{\begin{array}{c}  
    be \\ 
    ck \\
    gl \\
    dq \\
    hr \\
    ps 
   \end{array} \right\}\subset$ &
   $\left\{\begin{array}{c}  
    bem \\
    bet \\
    cfk \\
    ckt \\
    agl \\
    glt \\
    dfq \\
    dmq \\
    ahr \\
    hmr \\
    aps \\
    fps \\
       \end{array} \right\}\subset$  & 
   $\left\{ \begin{array}{c}  
    bemt \\
    cfkt \\
    aglt \\
    dfmq \\
    ahmr \\
    afps
    \end{array} \right\}+$ 
    
      $\left\{ \begin{array}{c}  
   chlq\\bgpq\\dgkr\\cepr\\bhks\\dels
    \end{array} \right\}$

     \medskip \\
 \end{tabular} 
\label{tab:comflow}
\end{equation*}
\end{center} 
\vspace{.2in}

With the $4 \times 4$ mutualistic system, there are six tipping cycles in $\hat{a_2}$, and as with the competitive case, each of these are in two tipping cycles in $\hat{a_1}$. However, unlike in the competitive case, there are eight new tipping cycles that have none of the $\hat{a_2}$ as subsets. In the next tipping cycle set, $\hat{a_0}$, it follows the compeitive case in having six sets, each of which consists of two sets from $\hat{a_1}$, and six sets that are new configurations (the same new configurations as in the competitive case). In addition, there are eight cycles each of which has as a subset one of the eight new configurations in $\hat{a_1}$. These eight cycles begin what might be called a parallel inclusion pattern, complicating the flow yet further:

\vspace{.2in}
\begin{adjustwidth}{-.8cm}{}
\begin{center}
\begin{equation*}
\begin{tabular}{l c c c}
    & $\hat{a_2}$ & $\hat{a_1}$  & $\hat{a_0}$ \\  
     \\
    & 
    $\left\{\begin{array}{c}  
    be \\ 
    ck \\
    gl \\
    dq \\
    hr \\
    ps 
   \end{array} \right\}\subset$ &
   $\left\{\begin{array}{c}  
    bem \\
    bet \\
    cfk \\
    ckt \\
    agl \\
    glt \\
    dfq \\
    dmq \\
    ahr \\
    hmr \\
    aps \\
    fps \\
    \end{array} \right\}+$  
    $\left\{ \begin{array}{c}  
     bgk\\cel\\bhq\\cpq\\der\\gpr\\dks\\hls
    \end{array} \right\}\subset$ &
 
   $\left\{ \begin{array}{c}  
    bemt \\
    cfkt \\
    aglt \\
    dfmq \\
    ahmr \\
    afps
    \end{array} \right\}+$ 
    
      $\left\{ \begin{array}{c}  
   bgkt\\celt\\bhmq\\cfpq\\demr\\agpr\\dfks\\ahls
    \end{array} \right\}+$

   $\left\{ \begin{array}{c}  
   chlq\\bgpq\\dgkr\\cepr\\bhks\\dels
    \end{array} \right\}$

     \medskip \\
 \end{tabular} 
\label{tab:mutflow}
\end{equation*}
\end{center} 
\end{adjustwidth}{}
\vspace{.2in}

Thus, the differences in the specific sign structure of the community matrices alters the flow of the inclusion properties of the tipping cycle sets as their cycle size increases. A full combinatorial characterisation of the flows in general yields the values reflected in the $\tilde{a_i}$ of the different forms, yet in a more intricate way.

For example, in the competitive case, the way each tipping cycle set unfolds as $i$ decreases for a given $n$ is as follows. For $n=8$, the first $\tilde{a}_i$ where $\tilde{a_i} > 0$ is $|\hat{a}_6|$,  

\begin{equation*}
|\hat{a}_6| = \frac{1}{2}\times\frac{8!}{6!}
\end{equation*}
The sets from $|\hat{a}_6|$ are then carried, through inclusion, as subsets into $|\hat{a}_5|$,
\begin{equation*}
|\hat{a}_{5_{\hat{a}_6}}| = 6 \times \frac{|\hat{a}_6|}{1}
\end{equation*}
where the expression $|\hat{a}_{5_{\hat{a}_6}}|$ indicates the sets first seen in $|\hat{a}_6|$ now included in the $|\hat{a}_5|$ sets, and where the value of $|\hat{a}_6|$ is divided by $1$ (the first step in which the set is included), and multiplied by $6$ (i.e. $i + 1$, the coefficient index of the smaller (in set size), preceding set). 

In the case of competitive systems, we know that the number of tipping cycle sets for each $a_i$ is $\frac{1}{2}\times\frac{n!}{i!}$, which for $i=5$ ($n=8$) equals the value of $|\hat{a}_{5_{\hat{a}_6}}|$. Therefore there are no new additional sets arising in $|\hat{a}_5|$ that are not supersets of $|\hat{a}_6|$.

The next coefficient consists of 

\begin{equation*}
|\hat{a}_{4_{\hat{a}_6}}| = 5 \times \frac{|\hat{a}_{5_{\hat{a}_6}}|}{2}
\end{equation*}
again, with $|\hat{a}_{5_{\hat{a}_6}}|$ divided by $2$ (the second step from $|\hat{a}_6|$), and multiplied by $5$ ($i+1$).

In addition to any supersets of $|\hat{a}_6|$ (via $|\hat{a}_{5_{\hat{a}_6}}|$), there are a number of newly formed sets ($|\hat{a}_{4_{\hat{a}_4}}|$) easily calculated from the total number of sets, 

\begin{equation*}
|\hat{a}_{4_{\hat{a}_4}}| = \frac{1}{2}\times\frac{8!}{4!} - |\hat{a}_{4_{\hat{a}_6}}|
\end{equation*}
Therefore, 
\begin{equation*}
|\hat{a}_4| = |\hat{a}_{4_{\hat{a}_6}}| + |\hat{a}_{4_{\hat{a}_4}}|.
\end{equation*}

This process can continue as $i$ decreases as follows, 

\begin{equation*}
|\hat{a}_{3_{\hat{a}_6}}| = 4 \times \frac{|\hat{a}_{4_{\hat{a}_6}}|}{3},
|\hat{a}_{3_{\hat{a}_4}}| = 4 \times \frac{|\hat{a}_{4_{\hat{a}_4}}|}{1}
\end{equation*}
where again the denominator indicates the number of steps away from the original set construction,
\begin{equation*}
|\hat{a}_{3_{\hat{a}_3}}| = \frac{1}{2}\times\frac{8!}{3!} - |\hat{a}_{3_{\hat{a}_6}}| - |\hat{a}_{3_{\hat{a}_4}}| 
\end{equation*}
and
\begin{equation*}
|\hat{a}_3| = |\hat{a}_{3_{\hat{a}_6}}| + |\hat{a}_{3_{\hat{a}_4}}| + |\hat{a}_{3_{\hat{a}_3}}|.
\end{equation*}

Continuing, 
\begin{equation*}
|\hat{a}_{2_{\hat{a}_6}}| = 3 \times \frac{|\hat{a}_{3_{\hat{a}_6}}|}{4},
|\hat{a}_{2_{\hat{a}_4}}| = 3 \times \frac{|\hat{a}_{3_{\hat{a}_4}}|}{2},
|\hat{a}_{2_{\hat{a}_3}}| = 3 \times \frac{|\hat{a}_{3_{\hat{a}_3}}|}{1}
\end{equation*}

\begin{equation*}
|\hat{a}_{2_{\hat{a}_2}}| = \frac{1}{2}\times\frac{8!}{2!} - |\hat{a}_{2_{\hat{a}_6}}| - |\hat{a}_{2_{\hat{a}_4}}| - |\hat{a}_{2_{\hat{a}_3}}|  
\end{equation*}

\begin{equation*}
|\hat{a}_2| = |\hat{a}_{2_{\hat{a}_6}}| + |\hat{a}_{2_{\hat{a}_4}}| + |\hat{a}_{2_{\hat{a}_3}}| + |\hat{a}_{2_{\hat{a}_2}}|  
\end{equation*}
and so on, down to $a_0$, 

\begin{equation*}
|\hat{a}_{1_{\hat{a}_6}}| = 2 \times \frac{|\hat{a}_{2_{\hat{a}_6}}|}{5},
|\hat{a}_{1_{\hat{a}_4}}| = 2 \times \frac{|\hat{a}_{2_{\hat{a}_4}}|}{3},
|\hat{a}_{1_{\hat{a}_3}}| = 2 \times \frac{|\hat{a}_{2_{\hat{a}_3}}|}{2},
|\hat{a}_{1_{\hat{a}_2}}| = 2 \times \frac{|\hat{a}_{2_{\hat{a}_2}}|}{1}
\end{equation*}

\begin{equation*}
|\hat{a}_{1_{\hat{a}_1}}| = \frac{1}{2}\times\frac{8!}{1!} - |\hat{a}_{1_{\hat{a}_6}}| - |\hat{a}_{1_{\hat{a}_4}}| - |\hat{a}_{1_{\hat{a}_3}}| - |\hat{a}_{1_{\hat{a}_2}}|
\end{equation*}

\begin{equation*}
|\hat{a}_1| = |\hat{a}_{1_{\hat{a}_6}}| + |\hat{a}_{1_{\hat{a}_4}}| + |\hat{a}_{1_{\hat{a}_3}}| + |\hat{a}_{1_{\hat{a}_2}}|  + |\hat{a}_{1_{\hat{a}_1}}|
\end{equation*}

\begin{equation*}
|\hat{a}_{0_{\hat{a}_6}}| =  \frac{|\hat{a}_{1_{\hat{a}_6}}|}{6},
|\hat{a}_{0_{\hat{a}_4}}| =  \frac{|\hat{a}_{1_{\hat{a}_4}}|}{4},
|\hat{a}_{0_{\hat{a}_3}}| =  \frac{|\hat{a}_{1_{\hat{a}_3}}|}{3},
|\hat{a}_{0_{\hat{a}_2}}| =  \frac{|\hat{a}_{1_{\hat{a}_2}}|}{2},
|\hat{a}_{0_{\hat{a}_1}}| =  \frac{|\hat{a}_{1_{\hat{a}_1}}|}{1}
\end{equation*}

\begin{equation*}
|\hat{a}_{0_{\hat{a}_0}}| = \frac{1}{2}\times\frac{8!}{0!} - |\hat{a}_{1_{\hat{a}_6}}|
- |\hat{a}_{1_{\hat{a}_4}}|
- |\hat{a}_{1_{\hat{a}_3}}|
- |\hat{a}_{1_{\hat{a}_2}}|
- |\hat{a}_{1_{\hat{a}_1}}|
\end{equation*}

\begin{equation*}
|\hat{a}_0| = |\hat{a}_{0_{\hat{a}_6}}|
+|\hat{a}_{0_{\hat{a}_4}}|
+|\hat{a}_{0_{\hat{a}_3}}|
+|\hat{a}_{0_{\hat{a}_2}}|
+|\hat{a}_{0_{\hat{a}_1}}| + |\hat{a}_{0_{\hat{a}_0}}|
\end{equation*}\\

In general, for competitive systems, and a given $n$,

\begin{equation*}
|\hat{a}_{n-2}| =  \frac{1}{2}\frac{n!}{(n-2)!}
\end{equation*}
and for $i < n-2$, for individual supersets derived from smaller $\hat{a}_j$ sets, each smaller set is divided by the number of steps ($j-i$) removed from their original construction (at index $j$) and the whole expression multiplied by $i + 1$, 

\begin{equation*}
|\hat{a}_{i_{\hat{a}_j}}| = (i+1) \times
\frac{|\hat{a}_{{i+1}_{\hat{a}_j}}|}{j-i}.
\end{equation*}
Therefore, for a given index $i$ the supersets are 
\begin{equation*}
\Omega_i = (i+1) \times \sum_{j=i+1,k=1}^{j=n-2,k=n-2-i} \frac{|\hat{a}_{{i+1}_{\hat{a}_j}}|}{k}
\end{equation*}
while the number of newly formed sets in each $i$ that are not supersets are,
\begin{equation*}
\Gamma_i = \frac{1}{2}\frac{n!}{i!} - \Omega_i.
\end{equation*}
The $\Gamma_i$ values for competitive systems up to $n=12$ are shown in Table 3. Each $|\hat{a}_i|$, consisting of the cumulative supersets and the newly formed sets, is then    
\begin{equation*}
|\hat{a}_i| =  \Omega_i + \Gamma_i.
\end{equation*}
 
While the $\Omega_i$ follows the same patterning for mutualistic and predator-prey systems (one difference with these two forms is that $\Gamma_{n-3} \ne 0$, unlike with the competitive case), a natural question arises as to what a full characterisation of the $\Omega_i$ and $\Gamma_i$ for any $n \times n$ system with any sign structure might suggest. That is, whether there is a meaningful relation between the combinatorial structure of the destabilising sets and the general stability properties of systems based on their sign structure.

\begin{table}
%\begin{adjustwidth}{-.6cm}{}
\tiny
\begin{tabular}{c c c c c c c c c c c}
 & $n$  & $a_8$ & $a_7$ & $a_6$ & $a_5$ & $a_4$ & $a_3$ & $a_2$ & $a_1$ & $a_0$ \\
      & $4$ &&& && &  &  &  & 6 \\
      & $5$  &&&&& &  &  & 30 & 20 \\
      &$6$  &&&&&  &  & 90 & 120 & 135 \\
      &$7$ &&&&  &  & 210 & 420 & 945 & 924 \\
      &$8$  &&&&& 420 & 1120 & 3780 & 7392 & 7420 \\
      &$9$ & &&& 756 & 2520 & 11340 & 33264 & 66780 & 66744 \\
      &$10$ &&&1260& 5040 & 28350 & 110880 & 333900 & 667440 &  667485 \\
      &$11$ & &1980 & 9240 & 62370 & 304920 & 1224300 & 3670920 &7342335  & 7342280 \\
      &$12$ &2970&15840& 124740 & 731808 & 3672900& 14683680 &44054010& 88107360 & 88107426 \\
    
\end{tabular}
%\end{adjustwidth}{}
\normalfont
\caption{\textbf{Table 3} \textit{Values of $\Gamma_i$ for each $|\hat{a}_i|$ for competitive systems up to $n=12$. There is rich structure among the $\Gamma_i$. For example, the $a_0$ values ($\{6,20,135,924,...\}$) are the rencontres numbers with two fixed points; the first entries for each $n$ ($\{\alpha_4,\alpha_5,\alpha_6,\alpha_7,...\}=\{6,30,90,210,...\}$) are $n(n-1)(n-2)(n-3)/4$ (or $\alpha_n=\alpha_{n-1}+(n-1)(n-2)(n-3)$); the second set of values ($\{\beta_5,\beta_6,\beta_7,\beta_8,...\}=\{20,120,420,1120,...\}$) in each $n$ can be described as $\beta_n=\beta_{n-1}+ (n-1)(2\binom{n-2}{3}+(n-2)\binom{n-3}{2})$; the third set of values ($\{\delta_6,\delta_7,\delta_8,\delta_9,...\}=\{135,945,3780,11340,...\}$) is $\delta_n=\delta_{n-1}+ (n-1)(\binom{\binom{n-3}{2}}{2}+(n-2)\frac{(n-3)!}{(n-6)!})$.}}
\end{table}
\clearpage
\vspace{.3in}
\centerline {\bf References}
\medskip
[1] B. Brooks, The coefficients of the characteristic polynomial in terms of the eigenvalues and the elements of an $n \times n$ matrix, Appl. Math. Lett., 19:511–515, 2006.
\medskip

[2] W. Geary, M. Bode, T. Doherty, E. Fulton, D. Nimmo, A. Tulloch, V. Tulloch, and E. Ritchie, A guide to ecosystem models and their environmental applications, 	Nat. Ecol. Evol., 4(11):1459-1471, 2020.
\medskip

[3] A. Hurwitz, Über die bedingungen, unter welchen eine gleichung nur wurzeln mit negativen reellen theilen besitzt, Math. Ann., 46:273–284, 1895.
\medskip

[4] A. Lotka, Elements of physical biology, Williams and Wilkins, Baltimore, Maryland, 1925.
\medskip

[5] N. Obrechkoff, Sur un probl\`{e}me de Laguerre. C.R. Acd. Sci. Paris, 177:223-235, 1923.
\medskip

[6] E. Routh, Treatise on the Stability of a Given State of Motion: Particularly Steady Motion, Macmillan and Co., London, 1877.
\medskip

[7] N. Sloane, The on-line encyclopedia of integer sequences, https://oeis.org/.
\medskip

[8] V. Volterra, Variazioni e fluttuazioni del numero d’individui in specie animali conviventi, Mem. Acad. Lincei Roma., 2:31–113, 1926.
\end{document}